\documentclass[preprint,aps]{revtex4}
\usepackage{graphics}
\usepackage{graphicx}
\usepackage{color}

\newcommand{\BV}{\left(\begin{array}{c}}
\newcommand{\EV}{\end{array}\right)}
\newcommand{\BM}{\left(\begin{array}{cc}}

\newcommand{\bqn}{\begin{equation}}
\newcommand{\eqn}{\end{equation}}
\newcommand{\beqry}{\begin{eqnarray}}
\newcommand{\eeqry}{\end{eqnarray}}

\begin{document}

\title{Fundamental length from algebra}
\author{T.\ Goldman}
\affiliation{Theoretical Division \\
Los Alamos National Laboratory \\
Los Alamos, NM 87545 USA \\
and \\
Department of Physics and Astronomy \\
University of New Mexico \\
Albuquerque, NM 87501 USA}

\vspace*{-0.75in}
\begin{flushright}
\today\\
LA-UR-19-30737\\
{arXiv:19mm.nnnnn}
\end{flushright}
\vspace*{0.75in}

\begin{abstract}

Advances in physics have required the application of more and more 
sophisticated mathematics. I present arguments supporting the contention 
that the next advance beyond quantum field theory will require the application 
of a non-associative algebra. The principle observable effect  would be the 
appearance of a fundamental length. An experimental search for such an 
effect may be feasible. 


\end{abstract}

\maketitle

\section{Introduction} \label{sec:intro}

When physics and astronomy first became quantitative disciplines, the number 
system employed was simply the real numbers. From Galileo through Newton 
and well into the 18th century C.E., the algebra of real numbers continued to 
suffice. With the advent of abstract wave phenomena, the advantages of using 
the algebra of complex numbers to describe phases became apparent. Although 
the application of groups to crystallography implicitly introduced non-commutative 
algebras, they did not become essential until the advent of quantum mechanics 
and non-Abelian gauge symmetries. These advances continued to suffice for quantum 
field theory with the caveat of renormalization, along with the expansion of geometry 
to $3+1$-dimensional space-time.

Separately, the study of gravity led to the inclusion of Riemanian geometries with 
their own mathematical complexities. Attempts to make further advances have included 
non-associative and non-commutative geometries, graded Lie algebras, also referred 
to as supersymmetric algebras, and associated with that, even higher space-time 
dimensions in string theory.~\cite{NAGFT,NAGDST} My comments below do not have 
any apparent direct relation to these latter questions. 

Instead, my focus is on the relation between the algebras in mathematics and the question 
of dimensional quantities in physics. 

\section{A Fundamental Scale} \label{sec:disc}

Einstein's $3+1$-dimensional space-time of special relativity~\cite{Ein} related spatial 
dimension and time by the universal constant of the speed of light, $c$. With the advent 
of the action principle approach to classical mechanics pursued for symmetries by 
Noether~\cite{Noe}, and its application to quantum mechanics, Planck's universal 
constant~\cite{Pla} 
was recognized to be both related to a non-commutative algebra and to define a relation 
between momentum and spatial dimension. In effect, these advances reduced the variety 
of dimensionful physical quantities to just one, which may be taken (with the assistance 
of $c$) to be described equivalently as mass or length. 

I turn next to the renormalization problem of quantum field theory. Although now 
well-regarded and viewed as an essential component of understanding the scale 
dependence of physical effects and quantities, it was initially viewed (by Feynman 
and others) as some sort of trick to avoid a fundamental question. By considering the 
nature of the associator compared with the nature of the commutator, we get an inkling 
of what that issue may be: A fundamental length that is a universal constant.

\subsection{Commutator}

The well-known quantum commutator between the spatial location operator, $q$, and 
the momentum operator, $p$, relates length and momentum to the Planck constant, 
$\hbar$, viz. 
\bqn
[ q , p ]  = \imath \hbar
\eqn 
or in dimensional terms, thinking of $p$ as some multiple ($ \beta\gamma $ in special relativity) 
of mass ($m$) times the speed of light ($c$) affords the opportunity to reduce all quantities 
to the dimensions of length. 

In quantum field theory, the commutator of scalar fields at two points, or equivalently a 
scalar field ($\phi(x)$) and its canonical momentum/spatial derivative ($\Pi_{\phi}(y)$) is 
similar (where $x$ and $y$ are spatial 3-vectors at the common time, $t=0$):
\bqn
[ \phi(x)  , \Pi_{\phi}(y) ]_{t=0} = \imath \hbar \delta^3(x-y).
\eqn 
So consistency requires the dimension of a scalar field be that of mass or, by the $\hbar$ 
equivalence, an inverse length as $\hbar$ (modulo $c$) reconciles the cubic dimension of 
mass on the $lhs$ with the inverse dimensions of length provided by the $\delta$-function 
on the $rhs$. 

\subsection{Associator}

Similarly, if we consider the associator of three scalar fields, 
\bqn
\{\phi(x) , \phi(y) , \phi(z)\}_{t=0} = A \hbar \delta^3(x-y) \delta^3(y-z)
\eqn
where $A$ is some product of purely mathematical constants and a dimensionful universal 
constant, it follows that the dimension of $A$ must be that of the cube of a length, $\ell_{f}$, 
(or equivalently, the inverse of the cube of a mass). I propose that this defines a fundamental 
length. (The factor of $\hbar$ has been included to reconcile the dimensions of the fields and 
one of the $\delta$-functions in what seems like a natural manner; eliminating it does change 
the number of powers of mass in the dimensionful constant, which may be significant.) 

\section{Value of Fundamental Length} \label{sec:size}

A value for a fundamental (Planck) length has been conjectured to be related to $\hbar$, and $c$ 
and Newton's constant for gravity, $G_{N}$,  as 
\bqn
\ell_{P\ell} = \sqrt{\frac{ \hbar G_{N}}{c^3}}  = 1.6 \times 10^{-35} {\rm m} 
\eqn
based on the concept that the corresponding ``point'' mass occupies a space smaller than that 
defined by its Schwarzschild radius and so must be a ``black hole'' such that anything entering 
that volume could not re-emerge. (It is difficult to see how Hawking radiation could arise from a 
point-particle black hole as opposed to one created initially from multiple components.) A similar 
length scale has been suggested on the basis of reconciling gravity and quantum mechanics~\cite{Adler}, 
even in the absence of a well-defined theory of quantum gravity.

If we consider $G_{N}$ in terms of a (Planck) mass, 
\bqn
G_{N}  = \frac{ \hbar c}{m_{P\ell}^2} 
\eqn
(where $m_{P\ell} \sim  2.2 \times 10^{-5} {\rm g}$), this is reminiscent of the discussion of 
the classical electron radius, $r_e$,  which relates the electrostatic self-interaction energy of a homogeneous charge distribution to the relativistic 
mass-energy of the electron, $m_e c^2$, namely, 
\bqn
r_{e,cl} = \alpha \frac{\hbar c}{m_e c^2}
\eqn
where $\alpha$ is the fine structure constant of electromagnetism. However, this is much 
smaller than the quantum ``size" of an electron at rest, 
\bqn
r_{e,qm} = \frac{\hbar c}{m_e c^2}
\eqn
which does not include the factor of the fine structure constant, so that 
\bqn
r_{e,qm} >> r_{e,cl}
\eqn
which has long been viewed as resolving the physical conundrum. 

I suggest that, similarly, that the value of $\ell_{f}$ in the associator will satisfy 
\bqn
\ell_{f}  >> \ell_{P\ell}
\eqn
on the basis that there might well be something parallel to the fine structure constant included 
in $G_{N}$ in a complete theory of quantum gravity. 

However, there is as yet no data supporting any particular value. If the most massive particle 
known,  the top quark, is indeed a ``point'' particle without substructure related to constituents 
of higher mass and shorter distance scale internal structure, we can only conclude that 
\bqn
\ell_{f}  \stackrel{<}{\sim}  10^{-18} {\rm m} 
\eqn
which is still about 17 orders of magnitude larger than the Planck length. This bound may be 
reduced by an order of magnitude or two by experiments measuring top quark production or 
scattering and comparing with calculations that allow for deviations from point-like structure via 
a top quark form factor, if no deviation is observed within the experimental and theoretical 
uncertainties. I return to this issue in my conclusions.

\section{Physics} \label{sec:fys}

A mathematical motivation for the choice of a non-associative algebra and the associator is 
straight-forward: This is the last of the classical algebras available, so it merits theoretical 
investigation in its own right. 

However, there is a better physical motivation. The divergences in renormalizable quantum field 
theories that lead to the requirement of renormalization only develop with interactions, {\it i.e.}, 
where three or more fields meet at a point, not just two. A physical question to ask then is: What 
difference might there be if the order in which the fields are brought into contact differs: $ x 
\rightarrow y$ at $z \neq x,y$ and $ z \rightarrow x,y$ {\it vs.} $ y \rightarrow z$ at $x \neq y,z$ 
and $ x \rightarrow y,z$. (Time ordering may be essential to understanding this, which raises 
questions regarding avoiding frame dependence.) The difference in the various sequences of 
forming combinations is what is described mathematically by the associator. 

For a black hole, the ``no-hair" theorems ensure that there is no difference, {\it i.e.}, the associator 
must vanish. However, if $\ell_{f}  >  \ell_{P\ell}$, there may well be a difference that is physically 
relevant to field theory. Relativity complicates this all, somewhat, by making the size of separations 
frame dependent, but it is not unreasonable to consider that the center-of-momentum frame for 
three converging particles (from the three field operators) is the appropriate one in which to view 
the effects of $\ell_{f}$. 

\section{Experiment and Application} 

The possibility of a direct experimental search for effects of the Planck length has been suggested 
by Bekenstein~\cite{Bek1,Bek2}. The feasibility has been examined by Maclay {\it et al.}~\cite{Mil} 
with the conclusion that while extremely difficult, the experiment may not be impossible. If the 
suggestion in this paper is valid, there is a motivation for embarking upon the search for the 
effect proposed by Bekenstein at every distance scale from the current limit down to the Planck 
length itself. An intermediate discovery may await an intrepid experimentalist. 

If an experimental discovery is made to the effect that $\ell_{f}  >> \ell_{P\ell}$, there may even be a 
bonus available to the theory of gravity. If we define the corresponding fundamental mass
\bqn
m_{f} = \frac{\hbar}{\ell_{f} c}
\eqn
then we might be able to identify
\bqn
G_{N}  = \zeta \frac{ \hbar c}{m_{f}^2}  .
\eqn
Consistency requires that $\zeta  <<  1$ if $m_{f}  <<  m_{P\ell}$. This allows for a possibility 
of describing why gravity is so weak (compared to other forces) in terms of some kind of effective 
coupling constant, although such a view might be at odds with General Relativity. 

\subsection{Other Applications} 

The associator and fundamental length offer opportunities to answer additional fundamental 
questions that have eluded resolution to the present. 

These include a resolution of the question of the nature of the Coulomb potential for a ``point" 
charge. A fundamental length would be expected to blunt the divergence at zero distance. A related 
issue is the electromagnetic self energy of such a charge and its contribution to the masses of the 
so-called fundamental particles. The Standard Model currently describes those masses in terms of 
the vacuum expectation value of the Higgs' field, but electromagnetic corrections are not ruled out 
by any means. Other (weak and strong interaction) self-energy contributions might also become 
calculable in a reliable and well-defined formulation.  

In principle, this question should also apply to the gravitational self-energy of a ``point" particle.
Although quantum effects will spread the wave function of such a particle over a larger volume 
than that defined by the Schwarzschild radius for its mass, except for a particle of mass $m_{P\ell}$, 
the self-energy question appears in principle just as it does for the other fundamental interactions. 
This last point may, however, be obviated by Mottola's observations~\cite{EM} of the quantum 
anomaly in the conservation equation for the divergence of the gravitational current.

\section{Acknowledgments}
I thank Peter Milonni for useful conversations. This work was carried out in part under the auspices 
of the National Nuclear SecurityAdministration of the U.S. Department of Energy at Los Alamos 
National Laboratory under Contract No. DE-AC52-06NA25396.

\end{document}